\documentstyle[twocolumn,prl,tighten,aps,amstex,graphicx,times,floats]{revtex}

\begin{document}

\thispagestyle{empty}

\marginparwidth 1.cm
\setlength{\hoffset}{-1cm}
\newcommand{\mpar}[1]{{\marginpar{\hbadness10000%
                      \sloppy\hfuzz10pt\boldmath\bf\footnotesize#1}}%
                      \typeout{marginpar: #1}\ignorespaces}
\def\mda{\mpar{\hfil$\downarrow$\hfil}\ignorespaces}
\def\mua{\mpar{\hfil$\uparrow$\hfil}\ignorespaces}
\def\mla{\marginpar[\boldmath\hfil$\rightarrow$\hfil]%
                   {\boldmath\hfil$\leftarrow $\hfil}%
                    \typeout{marginpar: $\leftrightarrow$}\ignorespaces}

\def\ba{\begin{eqnarray}}
\def\ea{\end{eqnarray}}
\def\bq{\begin{equation}}
\def\eq{\end{equation}}

\renewcommand{\abstractname}{Abstract}
\renewcommand{\figurename}{Figure}
\renewcommand{\refname}{Bibliography}

\newcommand{\eg}{{\it e.g.}\;}
\newcommand{\ie}{{\it i.e.}\;}
\newcommand{\etal}{{\it et al.}\;}
\newcommand{\ibid}{{\it ibid.}\;}

\newcommand{\mx}{M_{\rm SUSY}}
\newcommand{\pt}{p_{\rm T}}
\newcommand{\et}{E_{\rm T}}
\newcommand{\del}{\varepsilon}
\newcommand{\sla}[1]{/\!\!\!#1}
\newcommand{\fb}{\;{\rm fb}}
\newcommand{\gev}{\;{\rm GeV}}
\newcommand{\tev}{\;{\rm TeV}}
\newcommand{\abi}{\;{\rm ab}^{-1}}
\newcommand{\fbi}{\;{\rm fb}^{-1}}

\preprint{
\font\fortssbx=cmssbx10 scaled \magstep2
\hbox to \hsize{
\hskip.5in \raise.1in\hbox{\fortssbx University of Wisconsin - Madison}
\hfill\vtop{\hbox{\bf UB-HET-02-03}
            \hbox{\bf MADPH-02-1271}
            \hbox{\bf FERMILAB-Pub-02/092-T}
            \hbox{hep-ph/0206024}
            \hbox{\today}                    } }
}

\title{ 
Measuring the Higgs Boson Self Coupling at the LHC 
and Finite Top Mass Matrix Elements
}

\author{
Ulrich Baur${}^1$,
Tilman Plehn${}^2$, and
David Rainwater${}^3$
} 

\address{ 
${}^1$
Dept. of Physics, State University of New York at Buffalo, Buffalo, 
NY 14260, USA \\
${}^2$
Dept. of Physics, University of Wisconsin, Madison, WI 53706, USA \\
${}^3$
Theory Group, Fermi National Accelerator Laboratory, Batavia, IL 60510, USA
} 

\maketitle 

\begin{abstract}
  Inclusive Standard Model Higgs boson pair production and subsequent
  decay to same-sign dileptons via weak gauge $W^\pm$ bosons at the
  CERN Large Hadron Collider has the capability to determine the Higgs
  boson self-coupling, $\lambda$. The large top quark mass limit is
  found not to be a good approximation for the signal if one wishes to
  utilize differential distributions in the analysis. We find that it
  should be possible at the LHC with design luminosity to establish
  that the Standard Model Higgs boson has a non-zero self-coupling and
  that $\lambda / \lambda_{SM}$ can be restricted to a range of 0--3.7
  at $95\%$ confidence level if its mass is between 150 and 200~GeV.
\end{abstract}

\vspace{0.2in}


The CERN Large Hadron Collider (LHC) is widely regarded as capable of
directly observing the agent responsible for electroweak symmetry
breaking and fermion mass generation. This is generally believed to be
a light Higgs boson with mass
$M_H<200$~GeV~\cite{lepewwg}. Furthermore, the LHC promises complete
coverage of Higgs decay scenarios~\cite{tdr+}, including general
parameterizations in the Minimal Supersymmetric Standard
Model~\cite{tdr+,wbf_ll}, invisible Higgs decays~\cite{wbf_inv}, and
possibly even Higgs boson decays to muons~\cite{Hmumu}.  This broad
capability was made possible largely by the addition of the weak boson
fusion production channel to the search
strategies~\cite{wbf_ll,wbf_ww}. Observation of a Higgs boson in this
channel also contains additional information in the angular
distributions of the scattered quarks which reveal the fundamental
tensor structure of the $VVH$ vertex~\cite{vertex}.  With mild
theoretical assumptions, the Higgs boson total width, $\Gamma_H$, can
be determined via combination of all available channels, which in turn
yields the gauge and various Yukawa couplings~\cite{Hcoup,Yt}. The
weak boson fusion channels have received considerable attention in the
LHC experimental collaborations, and a number of detailed detector
simulation studies on them have already been completed, with very
encouraging results~\cite{wbf_exp}.

While these studies have shown that the LHC promises broad and
significant capability to measure various properties of the Higgs
sector, what remains is to determine the actual Higgs potential. This
appears in the Lagrangian as
\begin{equation*}
\label{eq:Hpot}
V(\Phi) \, = \, 
-\lambda v^2 (\Phi^\dagger\Phi) \, + \, \lambda (\Phi^\dagger\Phi)^2,
\end{equation*}
where $\Phi$ is the Higgs field, $v=(\sqrt{2}G_F)^{-1/2}$ is the
vacuum expectation value, and $G_F$ is the Fermi constant. In the
Standard Model (SM), $\lambda=\lambda_{SM}=M_H^2/(2v^2)$.  Regarding
the SM as an effective theory, $\lambda$ is {\it per se} a free
parameter. Its upper limit can be determined using unitarity
arguments, assuming the model's validity to high energy
scales~\cite{unit}. To measure $\lambda$, and thus determine the Higgs
potential, at a minimum experiments must observe Higgs boson pair
production; while this has been shown to be possible for a light Higgs
boson at a future Linear Collider~\cite{LC_HH}, no study has yet been
presented which suggests this is possible at the LHC for the SM Higgs
boson.

We simulate the signal process, pair production of two SM Higgs bosons
in gluon fusion, at the parton level for $pp$ collisions at $\sqrt{s}
= 14\tev$. Both Higgs bosons are decayed to $W$ boson pairs, which
subsequently are decayed to four jets and two same-sign
leptons~\footnote{While our study was in progress, Ref.~\cite{SLHC}
appeared. It includes a brief discussion of Higgs boson pair
production at an upgraded LHC, which would gather 20 times the amount
of data expected in the first run.}~\footnote{Unfortunately the search
in this channel cannot be generalized to the supersymmetric case,
since the branching fraction to $W$ bosons is suppressed for the light
Higgs scalar close to the decoupling regime.}:
\begin{equation*}
gg \to HH \to (W^+W^-) \; (W^+W^-)
          \to (jj \ell^\pm \nu) \; (jj {\ell'}^\pm \nu),
\end{equation*}
where $\ell,\,\ell'$ are any combination of electrons or muons. The
intermediate Higgs and $W$ bosons are treated off-shell using finite
widths in the double pole approximation. We calculate the signal using
two methods: exact loop matrix elements~\cite{higgs_self} and the
infinite top quark mass limit. The latter, which is commonly used in
place of exact matrix elements to speed up the calculation, reproduces
the correct total cross section for $HH$ production to within $10\%$
to $30\%$ for Higgs masses between 140~GeV and $200\gev$. However, it
produces completely incorrect kinematic distributions.

Signal results are computed consistently to leading order QCD with the
top quark mass set to $m_t=175\gev$ and SM top Yukawa coupling, and
the renormalization and factorization scales are taken to be the Higgs
boson mass~\cite{higgs_self}. The effects of next-to-leading (NLO)
order QCD corrections are included by multiplying the differential
cross section by an overall factor $K=1.65$ ($K$-factor), as suggested
by Ref.~\cite{higgs_nlo} where the QCD corrections for $gg\to HH$ have
been computed in the large $m_t$ limit. The multiplicative effects of
NLO QCD corrections are not expected to depend on whether the signal
is calculated with exact matrix elements or in the infinite top quark
mass limit.

The SM backgrounds of interest are those that produce two same-sign
leptons and four well-separated jets which reconstruct in two pairs to
a window around the $W$ boson mass. The largest contribution
originates from $W^{\pm}W^+W^-jj$ production, followed by
$t\bar{t}W^\pm$ where one top quark decays leptonically, the other
hadronically, and neither $b$ quark jet is tagged. Other backgrounds,
which in sum contribute at the $<5\%$ level~\cite{SLHC}, are:
$t\bar{t}t\bar{t}$ production, where none of the $b$ quark jets are
tagged, and additional jets or leptons are not observed; $W^\pm Z
jjjj$ production with leptonic $Z$ decay (including off-shell photon
interference) where one lepton is not observed; and $t\bar{t}j$ events
where one $b$ quark decays semileptonically with good hadronic
isolation, and the other is not tagged. For this letter we consider
only the two dominant backgrounds; since the others enter in sum at
less than $5\%$ of the total contribution, they do not change our
results noticeably.

\begin{figure}[t] 
\begin{center}
\includegraphics[width=8.5cm]{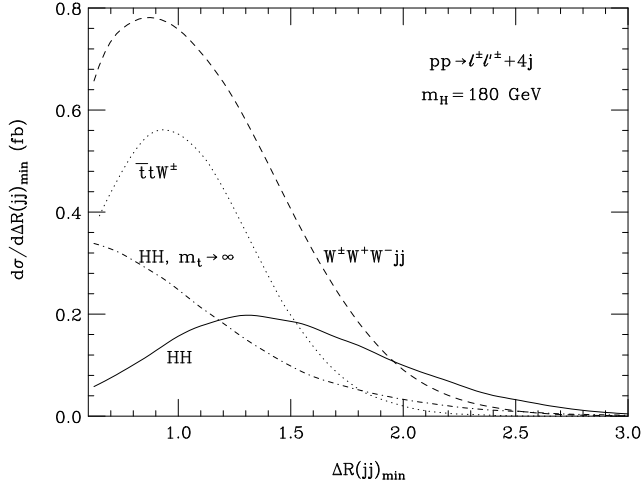}
\vspace*{2mm}
\caption[]{\label{fig:rjj} 
  Minimum separation between jets, $\Delta R(jj)_{\rm min}$, for the
  $M_H = 180\gev$, signal with exact matrix elements (solid line) and
  in the large $m_t$ limit (dot-dashed line), and the $WWWjj$ (dashed
  line) and $t\bar{t}W$ backgrounds (dotted line). Qualitatively
  similar results are obtained for other values of the Higgs boson
  mass in the range $150~{\rm GeV}\leq M_H\leq 200$~GeV.}
\vspace{-7mm}
\end{center}
\end{figure}

We simulate both leading backgrounds at the parton level using exact
matrix elements generated with {\sc madgraph}~\cite{madgraph}. For
$WWWjj$ production we evaluate the strong coupling constant $\alpha_s$
and the parton distribution functions at a scale $\mu$ given by
$\mu^2=\sum{p_T^2}$, where the sum extends over all final state
particles; for $t\bar{t}W$ production we take $\mu=m_t+M_W/2$. We use
a value for the strong coupling constant of $\alpha_s(M_Z) =
0.1185$. QCD corrections are not taken into account in our calculation
of $WWWjj$ and $t\bar{t}W$ production. The top quarks are generated on
shell (narrow width approximation), while all $W$ bosons in both
processes are allowed to be off shell. Assuming a $b$ quark tagging
efficiency of $50\%$, only 1/4 of the $t\bar{t}W$ rate contributes to
the background; events with one or two tagged $b$ quarks are
rejected. All signal and background cross sections are calculated
using CTEQ4L~\cite{cteq} parton distribution functions.

The kinematic acceptance cuts for both signal and backgrounds are:
\begin{eqnarray*}
&p_T(j) > 30,30,20,20 \gev , \qquad 
p_T(\ell) > 15,10 \gev    ,        \\
&|\eta(j)| < 3.0     ,        \qquad \qquad \qquad \qquad 
|\eta(\ell)| < 2.5   ,              \\
&\Delta R(jj) > 0.6 ,   \qquad 
\Delta R(j \ell) > 0.4 , \qquad 
\Delta R(\ell \ell) > 0.2 ,
\end{eqnarray*}
where $\Delta R =
\left[\left(\Delta\phi\right)^2+\left(\Delta\eta\right)^2\right]^{1/2}$
is the separation in the pseudorapidity -- azimuthal angle plane.  In
addition we require the four jets to combine into two pseudo-$W$ pairs
with invariant masses between 50~and $110\gev$ and assume that this
captures $100\%$ of the signal and backgrounds. We do not impose a
missing transverse momentum cut which would remove a considerable
fraction of the signal events.

\begin{figure}[t] 
\begin{center}
\includegraphics[width=8.5cm]{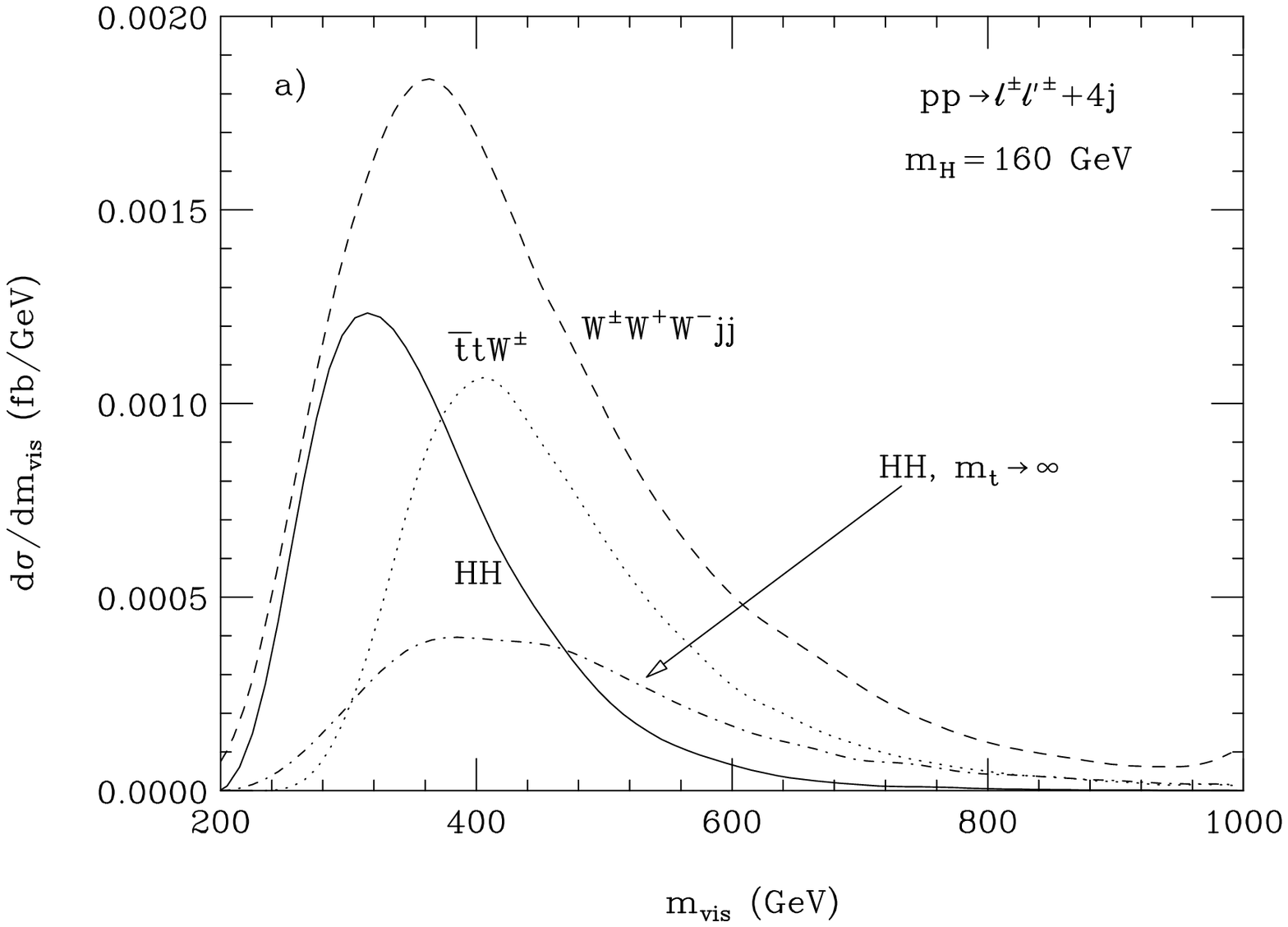} \\[3mm]
\includegraphics[width=8.5cm]{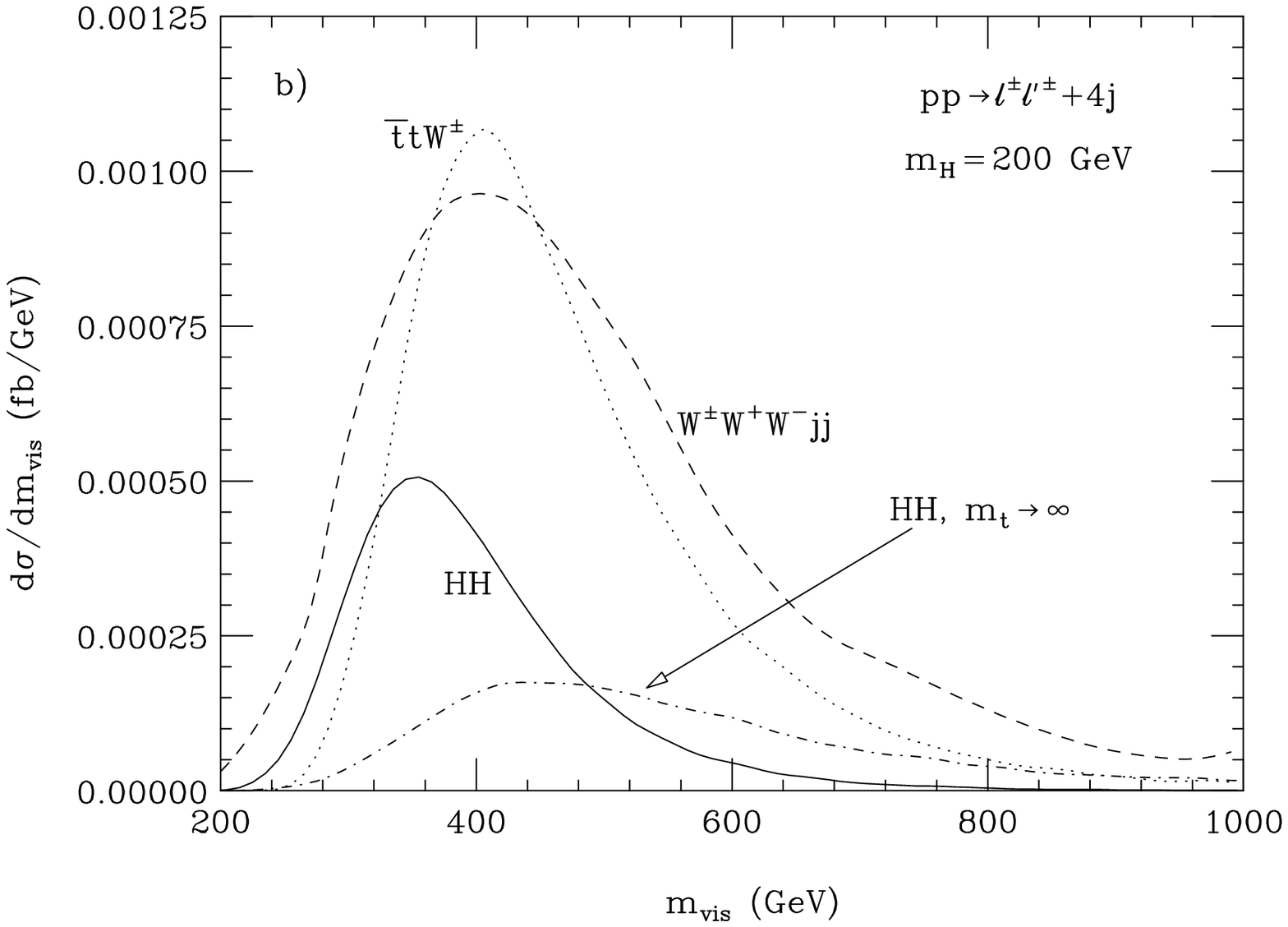}
\vspace*{2mm}
\caption[]{\label{fig:mvis} 
  Distribution of the invariant mass of all observable final state
  particles, $m_{vis}$, after all cuts, for the signal with a)
  $M_H=160\gev$ and b) $M_H=200\gev$, and the dominant
  backgrounds. The $m_{vis}$ distribution of the signal evaluated in
  the large $m_t$ limit is also shown.}
\vspace{-7mm}
\end{center}
\end{figure}

Both backgrounds are multi body production processes, and one
therefore expects that the distribution of the invariant mass,
$\sqrt{\hat s}$, of the system peaks at values significantly above
threshold. In contrast, the signal is a two-body production process
for which the $\sqrt{\hat s}$ distribution will exhibit a sharper
threshold behavior. Since the Higgs bosons are produced almost at
rest, the final state particles are distributed fairly isotropically,
resulting in a distribution of the minimum jet-jet separation, $\Delta
R(jj)_{\rm min}$, which is peaked at $\Delta R(jj)_{\rm min}\approx
1.3$ (see Fig.~\ref{fig:rjj}).  In contrast, the $\Delta R(jj)_{\rm
min}$ distribution for the background processes peaks at at a lower
value ($\Delta R(jj)_{\rm min}\approx 0.9$). In the following, we
therefore impose a more restrictive jet-jet separation cut of $\Delta
R(jj) > 1.0$, which results in a $\approx45\%$ reduction of the
background cross sections while reducing the signal only by about
$7-8\%$. Note that in the large $m_t$ limit the $\Delta R(jj)_{\rm
min}$ distribution of the signal peaks at $\Delta R(jj)_{\rm min}=0$
and drastically differs in shape from that calculated using the exact
loop matrix elements. If one were to calculate the signal cross
section in the large $m_t$ limit, a $\Delta R(jj)$ cut would not
result in a reduction of the background.

Unfortunately, with two neutrinos present in the final state,
$\sqrt{\hat s}$ cannot be reconstructed. However, we anticipate that
the invariant mass of all observed final state leptons and jets,
$m_{vis}$, will retain most of the expected behavior of the different
production processes. Figure~\ref{fig:mvis} clearly demonstrates that
this is the case: the signal peaks at lower values of $m_{vis}$ than
either background, especially for lower Higgs boson masses. However,
the $WWWjj$ background has a significant contribution from $WH(\to
W^+W^-)jj$ production, resulting in a $m_{vis}$ distribution which is
similar in shape to that of the $HH$ signal. Whereas the signal is
concentrated in the region $m_{vis}<500$~GeV, the background processes
have a significant tail extending to $m_{vis}=1$~TeV. This makes it
possible to normalize the background using data from the
$m_{vis}>500$~GeV region. Using exact loop matrix elements, the signal
displays a pronounced peak which gradually moves to higher values of
$m_{vis}$ with increasing Higgs boson mass. In contrast, in the large
$m_t$ limit, the $m_{vis}$ distribution of the signal is extremely
broad.

\begin{figure}[t] 
\begin{center}
\includegraphics[width=8.5cm]{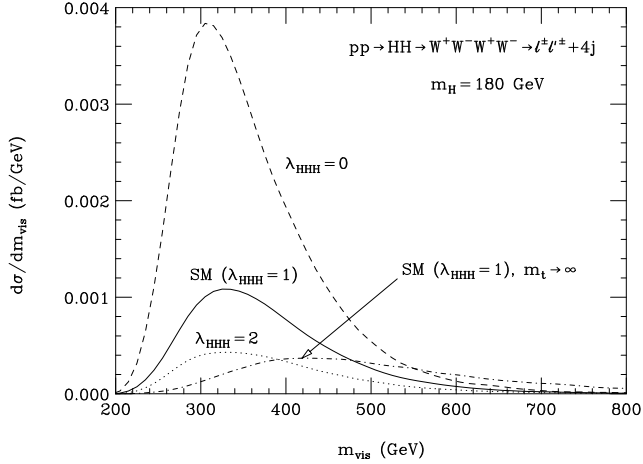}
\vspace*{2mm}
\caption[]{\label{fig:mvis_2} 
  The $m_{vis}$ distribution of the signal for $M_H=180$~GeV in the SM
  (solid curve), for $\lambda_{HHH}=\lambda/\lambda_{SM}=0$ (dashed
  line) and for $\lambda_{HHH}=2$ (dotted line). The dot-dashed line
  shows the SM cross section in the large $m_t$ limit. Qualitatively
  similar results are obtained for other values of $M_H$.}
\vspace{-7mm}
\end{center}
\end{figure}

The Feynman diagrams contributing to $gg\to HH$ in the SM consist of
fermion triangle and box diagrams~\cite{higgs_self}.
Non-standard Higgs boson self-couplings only affect the triangle
diagrams with a Higgs boson exchanged in the $s$-channel. They only
contribute to the $J=0$ partial wave, and thus impact the $m_{vis}$
distribution mostly at small values. This is illustrated in
Fig.~\ref{fig:mvis_2} for $M_H=180$~GeV and two values of
$\lambda_{HHH}=\lambda/\lambda_{SM}$. Since box and triangle diagrams
interfere destructively, the $gg\to HH$ cross section for
$1<\lambda_{HHH}<2.7$ is smaller than in the SM. The absence of a
Higgs boson self-coupling ($\lambda_{HHH}=0$) results in a Higgs pair 
production cross section which is about a factor~3 larger than the SM
result. 

\begin{figure}[t] 
\begin{center}
\includegraphics[width=8.5cm]{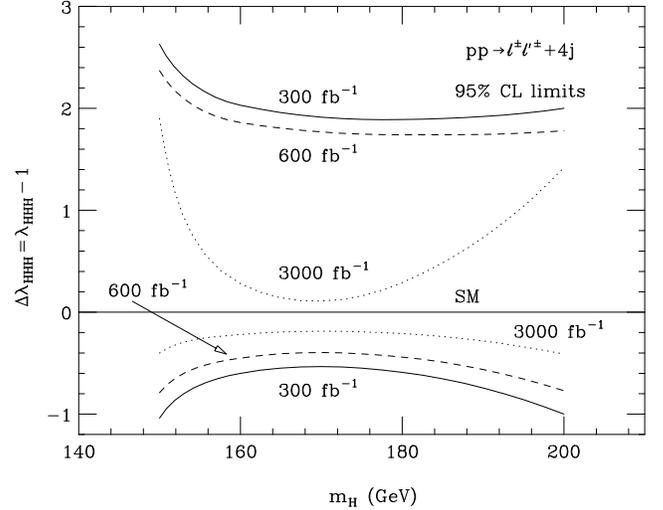}
\vspace*{2mm}
\caption[]{\label{fig:chisq} 
  Limits achievable at $95\%$ CL for
  $\Delta\lambda_{HHH}=\lambda_{HHH}-1$
  ($\lambda_{HHH}=\lambda/\lambda_{SM}$) in
  $pp\to\ell^\pm{\ell'}^\pm+4j$ at the LHC. Bounds are shown for
  integrated luminosities of 300~fb$^{-1}$ (solid lines),
  600~fb$^{-1}$ (dashed lines) and 3000~fb$^{-1}$ (dotted lines). The
  allowed region is between the two lines of equal texture.  The Higgs
  boson self-coupling vanishes for $\Delta\lambda_{HHH}=-1$.}
\vspace{-7mm}
\end{center}
\end{figure}

The shape change of the $m_{vis}$ distribution induced by non-standard
values of $\lambda_{HHH}$ can be used to derive quantitative
sensitivity bounds on the Higgs boson
self-coupling. Figure~\ref{fig:chisq} shows the $95\%$ confidence
level (CL) limits for $\Delta\lambda_{HHH}=\lambda_{HHH}-1$ which are
obtained from a $\chi^2$ test of the $m_{vis}$ distribution. The
allowed region is between the two lines of equal texture. In deriving
the bounds displayed, we combine channels with electrons and muons in
the final state, conservatively assuming a common lepton
identification efficiency of $\epsilon=0.85$ for each lepton. In order
to approximately take into account the (small) contributions to the
background from $t\bar tt\bar t$, $WZ+4j$ and $t\bar tj$ production
which we have ignored in our analysis, we scale the background
differential cross section by a factor~1.1. As mentioned before, our
calculation of the background processes does not include QCD
corrections which are expected to modify the relevant cross sections
by $20-40\%$. In order to derive realistic limits, we therefore allow
for a normalization uncertainty of $30\%$ of the SM cross
section. Since the background cross section can be directly determined
from the high $m_{vis}$ region with a statistical precision of $15\%$
or better for the assumed integrated luminosities, the bounds we
derive are conservative.

We derive sensitivity limits for integrated luminosities of
300~fb$^{-1}$, 600~fb$^{-1}$ and 3000~fb$^{-1}$, and Higgs boson
masses in the range $150~{\rm GeV}\leq M_H\leq 200$~GeV. Outside this
range, the number of signal events is too small to yield meaningful
bounds. For $M_H<150$~GeV, this is due to the small $H\to W^*W$
branching ratio. For $M_H>200$~GeV, the $gg\to HH$ cross section is
too small. An integrated luminosity of 300~fb$^{-1}$ (600~fb$^{-1}$)
corresponds to 3~years of running at the LHC design luminosity with
one (two) detectors. The larger value of 3000~fb$^{-1}$ can be
achieved in about 3~years of running if the planned luminosity upgrade
to ${\cal L}=10^{35}\,{\rm cm^{-2}\,s^{-1}}$~\cite{SLHC} is
realized. Figure~\ref{fig:chisq} demonstrates that, for 300~fb$^{-1}$,
a vanishing Higgs boson self-coupling ($\Delta\lambda_{HHH}=-1$) is
excluded at the $95\%$ CL or better, and that $\lambda$ can be
determined with a precision of up to $60\%$. Doubling the integrated
luminosity to 600~fb$^{-1}$ improves the sensitivity by $10-25\%$. For
300~fb$^{-1}$ and 600~fb$^{-1}$, the bounds for positive values of
$\Delta\lambda_{HHH}$ are significantly weaker than those for
$\Delta\lambda_{HHH}<0$, due to the limited number of signal events in
this region of parameter space. For 3000~fb$^{-1}$, the Higgs boson
self-coupling can be determined with an accuracy of $20-30\%$ for
$160~{\rm GeV}\leq M_H\leq 180$~GeV.

\medskip

\underline{In summary}, inclusive pair production of Higgs bosons at
the LHC, with decays to a same-sign lepton pair and four jets via four
$W$ bosons, will make it possible to perform a first, albeit not very
precise, measurement of the Higgs boson self-coupling $\lambda$.  The
non-vanishing of $\lambda$ can be established at $95\%$~CL or better
for $150~{\rm GeV}\leq M_H\leq 200$~GeV. The bounds on $\lambda$
derived here should be viewed as approximate. They can probably be
strengthened by including other final states such as $3\ell+2j$, or by
using more powerful statistical tools than the simple $\chi^2$ test we
performed. More details of our analysis, along with inclusion of
additional channels, will be presented elsewhere~\cite{3l-HH}.


\acknowledgements
We would like to thank S.~Dittmaier, M.~Spira and D.~Zeppenfeld for
useful discussions. One of us (U.B.) would like to thank the
Phenomenology Institute of the University of Wisconsin, Madison, and the
Fermilab Theory Group, where part of this work was carried out, for
their generous hospitality and for financial support.
This research was supported in part by the University of Wisconsin
Research Committee with funds granted by the Wisconsin Alumni Research
Foundation, by the U.~S.~Department of Energy under
Contracts No.~DE-FG02-95ER40896 and No.~DE-AC02-76CH03000, and the
National Science Foundation under grant No.~PHY-9970703.


\bibliographystyle{plain}

\end{document}